\documentclass[oneside,twocolumn,aps,prl,preprintnumbers,
superscriptaddress,amsmath,amssymb]{revtex4-2}

\usepackage{graphicx}
\usepackage{caption}
\usepackage{float}
\usepackage{color}
\usepackage{epstopdf}
\usepackage{amsmath}
\usepackage{braket}
\usepackage{amsthm}
\usepackage{amssymb}
\usepackage{amsfonts}
\usepackage{txfonts}
\usepackage{mathtools}
\usepackage{bm}
\usepackage{hyperref}
\usepackage{lipsum}
\usepackage{natbib}
\usepackage{physics}
\usepackage{cancel}
\usepackage{lipsum}
\usepackage{tikz}
\usepackage{comment}
\usepackage[justification=raggedright,singlelinecheck=false]{caption}

\usepackage{letltxmacro}
\LetLtxMacro{\originaleqref}{\eqref}
\renewcommand{\eqref}{Eq.~\originaleqref}

\newcommand{\Idoperator}{\mathbb{I}}

\newcommand{\re}{{\rm Re}}
\newcommand{\im}{{\rm Im}}

\def \ket#1{\mathinner{|{#1}\rangle}}
\def \bra#1{\mathinner{\langle{#1}|}}
\def\braket#1{\mathinner{\langle{#1}\rangle}}

\newcommand{\hlambda}{\hat{\lambda}}
\newcommand{\heta}{\hat{\eta}}
\newcommand{\hx}{\hat{x}}
\newcommand{\hp}{\hat{p}}
\newcommand{\hU}{\hat{\mathcal{U}}}
\newcommand{\hrho}{\hat{\rho}}
\newcommand{\G}{\mathcal{G}}
\newcommand{\ha}{\hat{a}}
\newcommand{\hadg}{\hat{a}^\dagger}
\newcommand{\has}{\hat{a}_{\sigma}}
\newcommand{\hasdg}{\hat{a}^\dagger_{\sigma}}

\newcommand{\tphi}{\tilde{\varphi}}

\begin{document}

\title{Operational criterion for Wigner function negativity}

\author{Paolo Solinas}
\email{paolo.solinas@ge.infn.it}
\affiliation{Dipartimento di Fisica, Universit\`a di Genova, via Dodecaneso 33, I-16146, Genova, Italy.}
\affiliation{INFN - Sezione di Genova, via Dodecaneso 33, I-16146, Genova, Italy.}

\author{Beatrice Donelli}
\affiliation{Istituto Nazionale di Ottica del Consiglio Nazionale delle Ricerche (CNR-INO), Largo Enrico Fermi 6, I-50125 Firenze, Italy.}
\affiliation{European Laboratory for Non-linear Spectroscopy, Universit\'a di Firenze, I-50019 Sesto Fiorentino, Italy.}

\author{Stefano Gherardini}
\affiliation{Istituto Nazionale di Ottica del Consiglio Nazionale delle Ricerche (CNR-INO), Largo Enrico Fermi 6, I-50125 Firenze, Italy.}
\affiliation{European Laboratory for Non-linear Spectroscopy, Universit\'a di Firenze, I-50019 Sesto Fiorentino, Italy.}

\begin{abstract}
We introduce an operational criterion to identify Wigner function (WF) negativity for an arbitrary quantum state within the framework of quantum non-demolition measurements. This criterion corresponds to experimentally accessible schemes that enable a direct measurement of the WF, and establishes the coherent-state basis as a privileged basis for determining when the WF exhibits negative regions.
We show that the absence (presence) of coherent superpositions in the coherent-state basis provides direct information about the positivity (negativity) of the WF. In particular, the absence of such superpositions constitutes a sufficient condition for WF positivity. Although a general proof of necessity remains elusive, we demonstrate that this condition is also necessary in two relevant cases: Schr\"{o}dinger-cat states and higher-order cat states on a circle.
More precisely, for Schr\"{o}dinger-cat states we establish a necessary and sufficient condition for the positivity of the WF in full generality, whereas for high-order cat states on a circle we derive an analogous condition in the limit of a large number of densely packed coherent states.
\end{abstract}

\maketitle

Discriminating whether a system behaves classically or exhibits quantum features is of paramount importance, both for the foundations of quantum mechanics and for quantum technologies. From a foundational perspective, identifying when quantum behavior in a system vanishes allows us to understand whether it undergoes a quantum-to-classical transition~\cite{Arndt1999,Gerlich2011,Aspelmeyer2022,FuchsScienceAdvances2024,Pedalino2026}. From a technological point of view, quantum properties are essential for performing cryptographic and computational tasks that are not possible with classical systems~\cite{nielsen-chuang_book}.

The boundary between the quantum and classical worlds is defined by the emergence of non-classical behaviors. While inequalities such as those by Bell~\cite{Bell1964} and Leggett-Garg~\cite{Leggett1985,Emary_2014} are routinely used to identify nonlocality and violations of macrorealism, respectively, a primary hallmark of quantumness for a state is served by the negativity of the corresponding Wigner function (WF)~\cite{WignerPhysRev1932, Hillery1984}. The negativity of the WF derives from the attempt to infer the system's state using non-commuting observables---namely, position and momentum---that cannot be measured simultaneously~\cite{GherardiniTutorial}. 

A foundational result in determining when the WF presents negative regions was obtained by Hudson in 1974~\cite{hudson1974}. He showed that the WF of a pure quantum state is positive if and only if the quantum state is a coherent state, i.e., it is represented with a Gaussian wave function~\cite{Glauber1963, Sudarshan1963, Klauder1985}. In contrast, for more general (i.e., mixed) states, apart from partial results~\cite{Mandilara2009}, a clear framework to establish when the WF is negative is still missing.

In this article, we propose a new and operational approach to the problem based on Quantum Non-Demolition Measurements (QNDM)~\cite{ThornePRL1978,Braginsky1980,BraginskyBook,Caves1980,CavesRevModPhys,GuerlinNature2007, Deleglise2008,solinas2015fulldistribution,solinas2021,solinas2022,solinas2024,SolinasReview2026}. Within this framework, we show that the WF can be experimentally measured using two ancillary detectors at a single time, or by performing a sequential measurement of the position and momentum. 

Building on earlier~\cite{hudson1974} and more recent results~\cite{Diosi2000, Diosi2002, Brody2025, solinas2024}, we identify coherent states as a privileged set, determining that the off-diagonal elements of the system's density-matrix expressed in the coherent-state basis (CSB) provide direct information about the negativity of the WF. 

We prove that the absence of such off-diagonal elements, representing quantum coherence terms in the CSB, implies that the WF is positive. For two relevant cases in the literature, i.e., the Schr\"{o}dinger-cat state~\cite{Deleglise2008} and higher-order cat states on a circle~\cite{Sedov2020, Serafini2004, Vlastakis2013}, we also prove the necessity condition for the WF negativity. As a result, we can accurately predict the critical residual coherence between coherent states above which the WF has negative regions. Our criterion provides a predictive tool for current experimental platforms, such as cavity QED~\cite{GuerlinNature2007, Deleglise2008} and mesoscopic systems~\cite{Maisi2014, Lebedev2016}, while clarifying the fundamental constraints that non-commutativity imposes on phase-space measurements.

{\it Wigner function via QNDM scheme---}
The WF~\cite{WignerPhysRev1932, Hillery1984, hudson1974} for a one-dimensional quantum system is defined as 
\begin{equation}\label{eq:WF_general}
W(X,P) = \frac{1}{4 \pi^2} \int d\lambda~d\eta~ e^{ -i \lambda X} e^{ -i\eta P} \G_{\lambda, \eta}\,, 
\end{equation}
where $\hbar$ is set to 1 and $\G_{\lambda,\eta} = \Tr[e^{ i (\lambda \hx + \eta \hp) } \hrho]$ is the quasi-characteristic function associated with $W(X,P)$~\cite{hudson1974}.

Our first result is that the quasi-characteristic function $\G_{\lambda,\eta}$ and, thus, the WF, can be obtained by performing a QNDM scheme~\cite{solinas2015fulldistribution, solinas2021, solinas2022, solinas2024, SolinasReview2026} that uses two ancillary systems. To show this, consider a tripartite system composed of a quantum system $\mathcal{S}$ and two detectors $\mathcal{D}_1$ and $\mathcal{D}_2$ such that $\mathcal{S}_{\rm tot} = \mathcal{S} \otimes  \mathcal{D}_1 \otimes \mathcal{D}_2$.
We denote by $\hlambda$ and $\heta$ the operators acting on the degrees of freedom of $\mathcal{D}_1$ and $\mathcal{D}_2$ respectively, while $\hx$ and $\hp$ are the position and momentum operators of $\mathcal{S}$.
The system-detector coupling is generated through the unitary operator $\hU_{\lambda,\eta}  = \exp{ i (\hx \otimes \hlambda + \hp \otimes \heta) }$ where $\hx \otimes \hlambda \equiv \hx \otimes \hlambda \otimes \Idoperator$ and $\hp \otimes \heta \equiv \hp \otimes \Idoperator \otimes \heta$.
We consider an arbitrary initial state $\hrho$ for $\mathcal{S}$, $\hrho_1 = (1/\sqrt{M_1}) \sum_{\lambda, \lambda'} \ketbra{\lambda}{\lambda'}$ for $\mathcal{D}_1$ and $\hrho_2 = (1/\sqrt{M_2}) \sum_{\eta, \eta'} \ketbra{\eta}{\eta'}$ for $\mathcal{D}_2$, where $M_1$ and $M_2$ are appropriate normalization constants.
The total density matrix after the coupling is given by $ \hat{R}_f = \hU_{\lambda,\eta} \hat{R}_0 \hU^\dagger_{\lambda,\eta}$, where $\hat{R}_0= \hrho \otimes \hrho_1  \otimes \hrho_2$.

As in the QNDM scheme \cite{solinas2015fulldistribution, solinas2021, solinas2022, solinas2024, SolinasReview2026}, the quasi-characteristic function $\G_{\lambda,\eta}$ in \eqref{eq:WF_general} can be obtained from the phases accumulated in the two detectors between the states $\ket{\pm \lambda/2}$ and $\ket{\pm \eta/2}$~[see also the Appendix]. This fact becomes apparent by introducing the non-commuting, annihilation and creation operators $\ha = ( \hx/\sigma + i \sigma \hp )/\sqrt{2}$ and $\hadg = ( \hx/\sigma - i \sigma \hp)/\sqrt{2}$ with real $\sigma$, and, using the canonical commutation relation $[\hx, \hp ] = i$, we obtain:
\begin{eqnarray}
	\G_{\lambda, \eta} 
    = \Tr[e^{ \frac{i\lambda}{2} \hx} e^{\frac{i\eta}{2} \hp } \hrho \, e^{\frac{i\eta}{2}\   \hp } e^{ \frac{i\lambda}{2} \hx} ]
    = e^{-\frac{|\alpha|^2}{2}} \Trace \left[ e^{-\alpha^* \ha} \hrho \, e^{\alpha \hadg} \right],  
	\label{eq:G_def}
\end{eqnarray}
where $\alpha = \frac{ i \lambda \sigma^2 - \eta }{\sqrt{2} \sigma}$. The separation of $\hat p$ and $\hat x$ in the exponential allows us to interpret \eqref{eq:G_def} as given by two sequential non-demolition measurements within the QNDM scheme.

This link with QNDM suggests the following experimental protocol: the system is sequentially coupled in time to the two detectors $\mathcal{D}_1, \mathcal{D}_2$ so that information about the system's position and momentum is encoded in the detectors' phases. By varying the coupling strength $\lambda$ and $\eta$, the quasi-characteristic function $\G_{\lambda,\eta}$ can be reconstructed through standard interferometric measurements of the detector states. The Wigner function is then obtained via the inverse Fourier transform of $\mathcal{G}_{\lambda,\eta}$. In this framework~\cite{solinas2015fulldistribution,solinas2021,solinas2022,solinas2024}, the WF can be interpreted as the quasi-probability originated from the sequential measurement of the non-commuting operators $\hp$ and $\hx$. 

A more intriguing interpretation is suggested by \eqref{eq:G_def}. In the QNDM scheme, the $\G_{\lambda,\eta}$ can be seen as a tool to extract information about the non-Hermitian adjoint operators $\ha$ and $\ha^\dagger$, or, more precisely, about their expectation value w.r.t.~the arbitrary initial state $\hrho$. This description is formally equivalent to a sequential measurement of position and momentum of the system, where the action of the detectors on the system's state is described by a positive operator-valued measure (POVM); hence, no fundamental principle is contradicted~\cite{holevo1982probabilistic,kraus1983states,durr1992quantum, peres1993quantum}.

{\it Quantum coherence in the coherent-state basis---}
As discussed in Refs.~\cite{solinas2024,SolinasReview2026}, in a finite-dimensional space with an orthonormal basis, the negativity of the QNDM quasi-probability distribution arises from the presence of off-diagonal density-matrix elements in the eigenbasis of the measured observable. In the present case of an infinite-dimensional space, it is natural to ask to what extent a similar statement can be formulated. Considering the second term in \eqref{eq:G_def}, we identify the annihilation operator $\hat a$ as the measured observable and, consequently, the coherent states $|\beta,\sigma\rangle$ as the privileged basis, since $\hat a \ket{\beta,\sigma} = \alpha \ket{\beta,\sigma}$~\cite{Glauber1963, Sudarshan1963,Klauder1985}.

There is strong evidence that the CSB plays a privileged role in the description of the WF. First, Hudson’s theorem~\cite{hudson1974} affirms that a pure quantum state has a positive WF {\it if and only if} it is Gaussian, i.e., a coherent state. Second, Di\'osi and Kiefer~\cite{Diosi2000,Diosi2002}, as well as Brody et al.~\cite{Brody2025}, have shown that the pointer basis for the open-system dynamics of the WF is the Gaussian (coherent-state) basis. 

Since the CSB depends on a length scale $\sigma$ (i.e., the Gaussian width), one may further ask how $\sigma$ should be chosen when decomposing the system's density matrix over the CSB. It turns out that, although both the decomposition and the density matrix's off-diagonal elements depend on $\sigma$, the WF and its negativity do not. In fact, as discussed in Ref.~\cite{Diosi2002} and Appendix, the WF is invariant under the rescaling by $c$: $X \rightarrow c X$ and $P \rightarrow P/c$. Since this transformation corresponds to a change in $\sigma$, it follows that the quasi-characteristic function $\mathcal{G}_{\lambda,\eta}$ in \eqref{eq:G_def} can be decomposed in any CSB (which depends on $\hat a$ through $\sigma$). This invariance is implicit in Hudson’s theorem~\cite{hudson1974}.

In some physical situations, determining the length scale $\sigma$ is natural, for instance, in the case of a cat or higher-order cat state (see discussion below). Moreover, there exists a broad class of systems for which $\sigma$ is dictated by the physical problem itself, namely systems interacting with an environment, where the natural length scale is set by system-environment interactions~\cite{Diosi2000,Diosi2002,Brody2025}. In all these cases, it is convenient to use the CSB associated with this specific length scale.

More generally, the system may involve multiple length scales, as in the case of a superposition of two coherent states with different $\sigma$ values. However, the invariance of the WF under length rescaling ensures that any CSB can be used. In the following, to simplify the discussion, we restrict our analysis to cases in which the system is characterized by a single, known length scale $\sigma$.

{\it Sufficient condition for Wigner function negativity---} The absence of quantum coherence in the CSB $\ket{\beta,\sigma} \equiv \ket{\beta}$ constitutes a sufficient condition for the positivity of the WF. To see this, consider a system prepared in an incoherent mixture of coherent states. In this case, the density matrix is diagonal in the CSB and can be written as $\hrho_{\rm diag} = \int d^2 \beta  \braket{\beta|\hrho|\beta} \ketbra{\beta}{\beta}$ with $d^2 \beta = d \re \beta~d \im \beta$ and $\braket{\beta|\hrho|\beta}\geq 0$~\cite{Glauber1963}.
Then, the corresponding quasi-characteristic function reads $\G_{\lambda,\eta} = \int d^2 \beta  \braket{\beta|\hrho|\beta}$.
By integrating $\mathcal{G}_{\lambda,\eta}$ over the variables $\lambda, \eta$, one can straightforwardly show that the WF is positive; refer also to the Appendix for more details. This result naturally includes the sufficient condition of Hudson’s theorem~\cite{hudson1974}, corresponding to the case of a single coherent state, $\hat\rho = \ketbra{\beta}{\beta}$.

{\it Schr\"{o}dinger-cat states: Necessary condition for Wigner function negativity---}
The necessary condition is much more difficult to establish, as the CSB is infinite-dimensional and overcomplete \cite{Glauber1963, Sudarshan1963, Klauder1985}. However, for specific but important situations in which the system’s symmetries can be exploited, analytical results can be obtained. These relevant cases include the Schr\"{o}dinger-cat state and high-order cat states on a circle. To introduce them formally, we consider a system prepared in a superposition of coherent states~\cite{Sedov2020,Serafini2004,Vlastakis2013}: $\ket{\psi_0} = \sum_k \psi_{0,k} \ket{\beta_k}$, which corresponds to the density matrix
\begin{equation}
    \hat{\rho} = \frac{1}{N}\sum_{j,k} \, \rho_{j,k} |\beta_j\rangle\!\langle
    \beta_k|, 	
	\label{eq:discrete_rho}
\end{equation}   
where $N=\sum_{k,j}\rho_{j,k} \, e^{-\frac{1}{2}\left( |\beta_j|^2 + |\beta_k|^2 \right) + \beta_j \beta_k^*}$, and $\rho_{i,j}, \beta_i, \beta_j$ are complex numbers. Using \eqref{eq:WF_general}, we obtain (see Appendix): $W(X,P) = N^{-1}\sum_{j,k} \rho_{j,k} W_{j,k}(X, P)$, where 
\begin{align}\label{eq:WF_jk}
    W_{j,k}(X, P) &= \frac{1}{\pi} e^{ -\frac{1}{\sigma^2} \left(X-\frac{X_j+X_k}{2}\right)^2
    -\sigma^2 \left(P-\frac{P_j+P_k}{2}\right)^2 } \nonumber\\
    &\times  e^{ i \left[ X (P_j-P_k)-P (X_j-X_k)+\frac{(X_j P_k - X_k P_j)}{2 } 
    \right]}\,.    
\end{align}

\begin{figure}
\centering
\includegraphics[width=1.03\columnwidth]{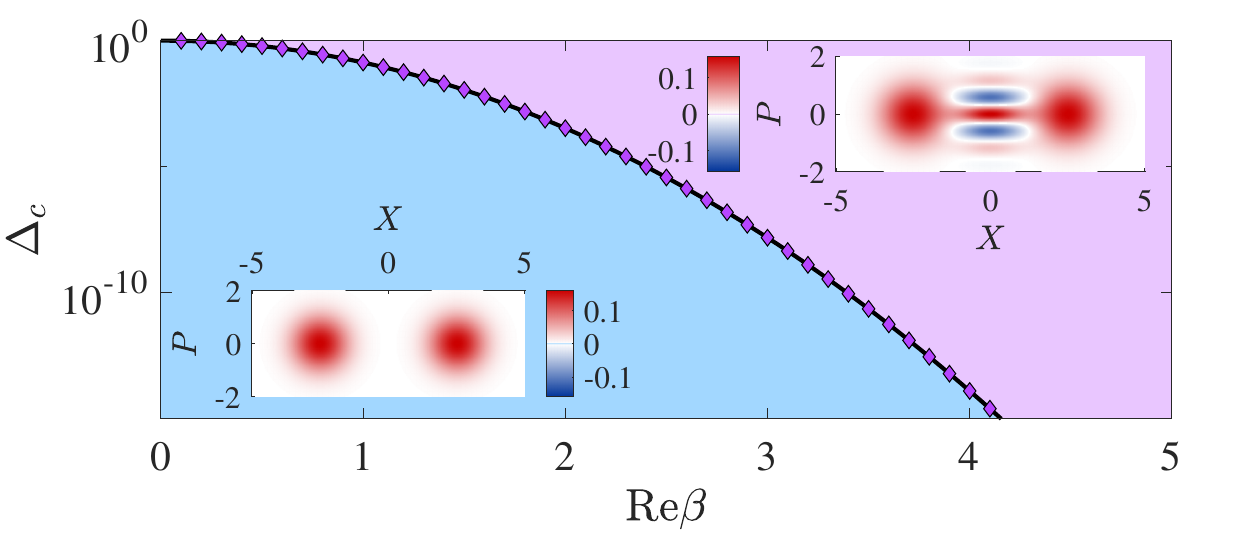}
\vspace{-15pt}
\caption{
Schr\"{o}dinger-cat state: critical residual coherence $\Delta_c$ as a function of the distance $\Re\beta$ of the coherent states from the origin. The analytical prediction of \eqref{eq:Delta_c} (black curve) separates the region in which the WF is positive (light blue) from the one where it is negative (violet). The purple diamonds represent the values of $\Delta_c$ obtained numerically.
In the insets, we plot the cat state with interference, i.e., $\Delta = 1$ (upper right), and without interference, i.e., $\Delta = 0$ (lower left), corresponding to the cases in which the WF is negative and positive, respectively. The coherent states are centered at $(\pm 2\sqrt{2}, 0)$, and all lengths are rescaled w.r.t.~$\sigma$. 
}
\label{fig:fig1}
\end{figure}

The Schr\"{o}dinger-cat state is obtained by restricting to $j,k = 1,2$ in Eqs.~(\ref{eq:discrete_rho})–(\ref{eq:WF_jk}), corresponding to a superposition of two coherent states. Any cat state can be mapped, by an appropriate rotation and translation, to a configuration in which the coherent states are located at $\{X_1, P_1\} = \{\sqrt{2} \,\Re \beta, 0\}$ and $\{X_2, P_2\} = \{-\sqrt{2}\,\Re \beta, 0\}$ (see Fig.~\ref{fig:fig1}). Since rotations and translations do not affect the positivity or negativity of the WF, we can restrict our analysis to this case.

Our aim is to establish the relation---in terms of a necessary condition---between the presence of coherent superpositions in an initial cat state $\hat{\rho}$ [\eqref{eq:discrete_rho}], and the negativity of the corresponding WF. The diagonal elements of $\hat{\rho}$ can be parametrized as $\rho_{1,1} = \cos^2(\theta)$ and $\rho_{2,2} = \sin^2(\theta)$, while the off-diagonal elements as $\rho_{1,2} = \Delta \cos(\theta) \sin(\theta) e^{i \phi}$ with $0 \leq \Delta \leq 1$~(see Appendix). 
The coherence parameter $\Delta$ enables us to relate the loss of coherence to the negativity of the WF that, for $j,k = 1,2$, takes the simplified expression~(see Appendix).
\begin{equation}\label{eq:WF_cat_state}
    W(X,P) = \frac{e^{-X^2- P^2 - 2\re \beta^2}}{ \pi N Z} 
    \left[ \cos^2\theta~Z^{2} + b\,Z + \sin^2\theta \right],
\end{equation}
where $Z = e^{2 \sqrt{2} X \re\beta }$, $b=e^{2 \re \beta^2} \Delta \big| \sin(2\theta) \big|\cos\tilde{\varphi}$ and $\tphi = \phi - 2 \sqrt{2} P \re \beta$. 

The zeros of the quadratic polynomial $\mathcal{P}(Z)=\cos^2\theta~ Z^{2} + b\,Z + \sin^2\theta$ in the WF are
\begin{equation}
    Z_{\pm} = \frac{- b \pm \sqrt{ b^2 - \sin^2(2\theta)}}{2 \cos^2 \theta};
    \label{eq:Z_equation}
\end{equation}
they identify the $(X,P)$-curves where $W(X, P)=0$. Since $\cos^2\theta >0$, the parabola $\mathcal{P}(Z)$ is concave up, and its minimum is at $Z_m = (Z_+ + Z_-)/2$. In particular,
\begin{equation}
    \mathcal{P}(Z_m) = \sin^2\theta \left(1- \Delta^2 \cos^2 \tphi~e^{4 \re \beta^2} \right). 
    \label{eq:P_min}
\end{equation}
As $(e^{-2 X^2- 2 P^2 - 2 \re \beta^2})/( \pi N Z)$ is non-negative, the condition to have a negative WF in \eqref{eq:WF_cat_state} is given by the inequality $\mathcal{P}(Z_m)<0$, which corresponds to $\Delta > e^{-2 \re \beta^2}/|\cos\tphi|$. The strongest bound is obtained when the cosine term is maximal, $\cos\tphi=1$, i.e., for $\tphi =2\pi k$ corresponding to momenta $P_{\rm max} = (\varphi - 2 k \pi)/(2 \sqrt{2} \, \re\beta^2)$. In this case, the condition for WF negativity reduces to
\begin{equation}
    \Delta > \Delta_c = e^{-2(\mathrm{Re}\,\beta)^2} \,.
    \label{eq:Delta_c}
\end{equation}

Equation~(\ref{eq:Delta_c}) is the main result of the paper. It implies that if the residual coherence exceeds the critical value $\Delta_c$, the WF is guaranteed to be negative. Moreover, the value of the minimum can be determined explicitly by solving Eqs.~(\ref{eq:Z_equation})-(\ref{eq:P_min}) under the condition $P = P_{\rm max}$. Therefore, \eqref{eq:Delta_c} provides a necessary and sufficient condition for the presence of negative regions in the WF of the Schr\"{o}dinger-cat states.

The behavior of $\Delta_c$ as a function of $\Re\beta$ is shown in Fig.~\ref{fig:fig1}. The black curve [\eqref{eq:Delta_c}] separates the parameter region in which the WF is positive from the one where it exhibits negativity. We verified that the theoretical prediction exactly matches the value of $\Delta$ obtained numerically (purple diamonds in Fig.~\ref{fig:fig1}).

It is worth highlighting the difference between the result of \eqref{eq:Delta_c} and those in \cite{solinas2024,SolinasReview2026}.
For states defined over a discrete and orthonormal basis, the presence of quantum coherence w.r.t.~the eigenstates decomposing the measured observable is enough to have a negative quasi-probability, in terms of the QNDM scheme. On the contrary, for Schr\"{o}dinger-cat states defined over a continuous and over-complete basis, we have a more stringent condition [\eqref{eq:Delta_c}] since the quantum coherence must exceed the critical value $\Delta_c$. However, we stress that, because $\Delta_c$ vanishes exponentially with $\re \beta$, in many practical cases, the discrete ~\cite{solinas2024,SolinasReview2026} and continuous basis cases coincide and the negativity of the WF can be directly related to the presence of coherent superpositions. The differences become important when the Gaussian functions forming the cat state are superposed, i.e., $\re\beta<1$.

\begin{figure}
\centering
\includegraphics[width=\columnwidth]{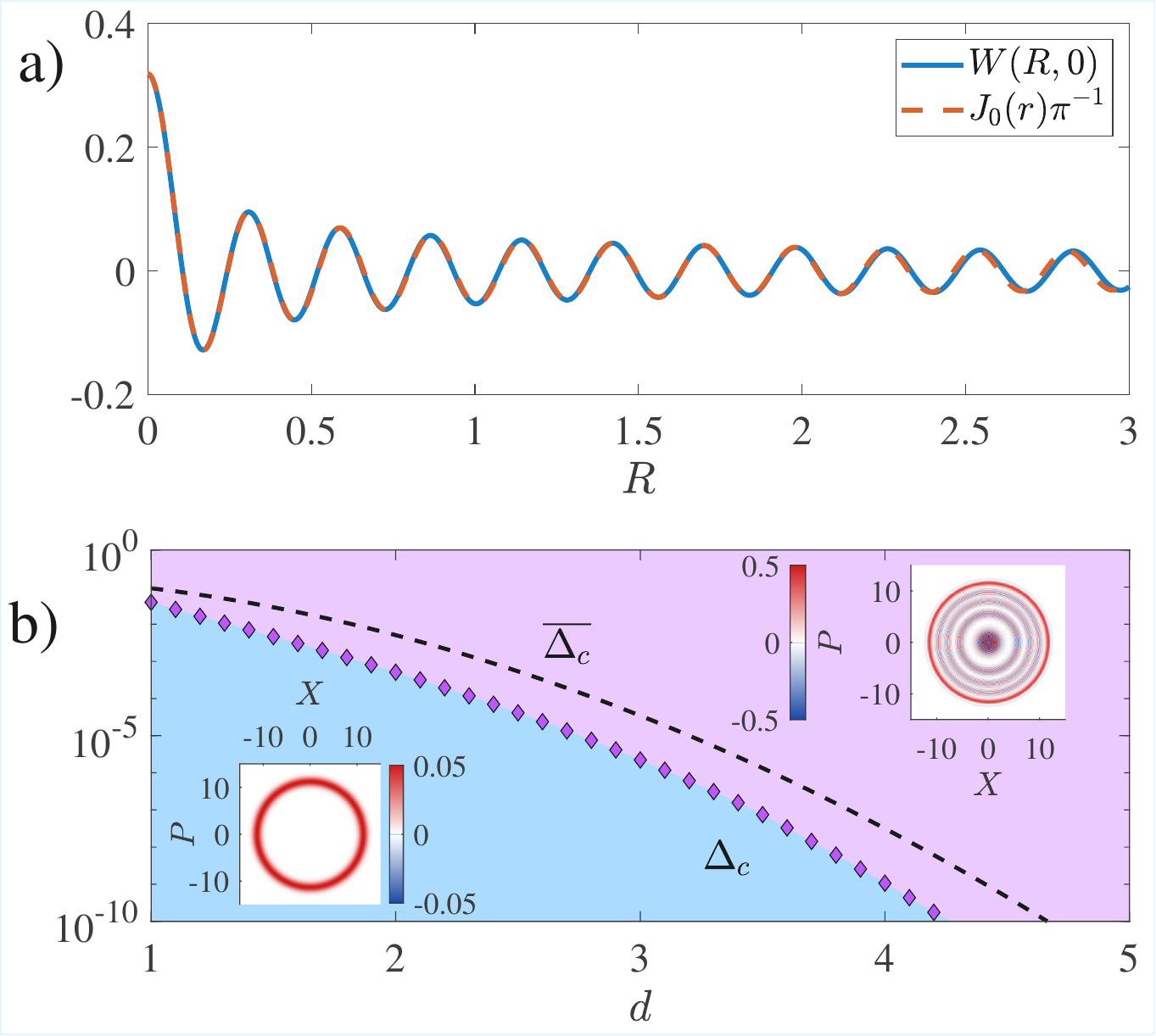}
\caption{
Higher-order cat states. 
a) Comparison of the radial dependence of the WF [\eqref{eq:WF_high_order_cat}] (solid blue curve) for $M=64$ coherent states arranged on a circle of radius $d=8$ obtained numerically for $\Delta=1$, with the theoretical prediction $J_0(r) / \pi$ (red dashed curve).
b) Critical residual coherence $\Delta_c$ as a function of $d$ (still $M=64$). The values of $\Delta_c$, determined numerically, are represented by the purple diamonds, and they separate the parameter regions in which the WF is positive (light blue) or negative (purple). The black dashed line is the upper bound $\overline{\Delta_c}$ given by \eqref{eq:delta_c_high_order}. 
In the insets, we plot the cat state with interference, i.e., $\Delta = 1$ (upper right), and without interference, i.e., $\Delta = 0$ (lower left), corresponding to the cases in which the WF is negative and positive, respectively. All lengths are rescaled w.r.t.~$\sigma$.
} 
\label{fig:fig2}
\end{figure}

{\it Higher-order cat states on a circle---}
The second relevant case-study we analyze is a quantum state comprising an {\it even} and a large number of coherent states $M$ equally populated and symmetrically distributed on a circle of radius $d$ (with $d$ real and positive). Let us thus discuss the conditions determining the negativity of the WF for such a packed cat state, with $M$ finite and $M \gg d^2$, where the interference patterns are generated inside the circle as shown in Fig.~\ref{fig:fig2} b). 

Given the symmetry of the state, we work with the polar coordinates $\{X, P \} = \{R \cos \phi, R \sin \phi \}$. The position of the coherent states in the $X{-}P$ space are $\{ X_k, P_k \} =  \sqrt{2}\{ d \cos \theta_k, d \sin \theta_k\}$ with $\theta_k = 2 \pi k/M$ and $1 \leq k \leq M$. To account for the dephasing process in the CSB~\cite{Brody2025, Diosi2000, Diosi2002}, we assume that $\rho_{j,k} = [(1-\Delta) \delta_{jk} +\Delta]/M$ with $0 \leq \Delta \leq 1$. As above, all the off-diagonal matrix elements decrease as a result of the dephasing, which we have taken to be uniform on all the coherent states. Hence, the WF from Eqs.~(\ref{eq:discrete_rho})-(\ref{eq:WF_jk}) reads as (see Appendix): 
\begin{eqnarray}\label{eq:WF_high_order_cat}
    W(R,\phi) &=& \frac{e^{-R^2-d^2}}{\pi N M} \sum_{j=1}^{M}\sum_{k=1}^{M} \left[ (1-\Delta)\delta_{jk} + \Delta \right] \nonumber \\ 
    &\times& \exp\left( -d^2 e^{2 i \Phi_{j,k}} + r \cos(\phi-\Lambda_{j,k}) e^{i\Phi_{j,k}} \right), 
\end{eqnarray}
where $r\equiv 2\sqrt{2}R d$, $\Lambda_{j,k}=\frac{\theta_j+\theta_k}{2}$ and $\Phi_{j,k}=\frac{\theta_j-\theta_k}{2}$. In the sums of \eqref{eq:WF_high_order_cat}, we separate the diagonal terms ($j=k$) from the off-diagonal ones ($j \neq k$), such that $W(R, \phi) = W_{D}(R, \phi) + W_{OD}(R, \phi)$. Setting $\ell \equiv j-k$, we have $\Phi_\ell=\pi \ell/M$ and $\Lambda_{\ell,k}=(\ell +2k)\pi/M$. The two contributions are then evaluated using the Jacobi-Anger expansion~\cite{abramowitz+stegun} (see Appendix).

We look for the conditions ensuring the negativity of the WF inside the circle of radius $R$ (see Fig.~\ref{fig:fig2}). Notice that, as the length of all the quantities is rescaled in order to have $\sigma=1$, the results we are now going to show are valid for $d\geq 1$. Close to the origin, i.e.~$R {<} M/(2\sqrt{2}d)$, and in the limit of highly packed states ($M$ finite and $M \gg d^2$), the dominant contributions of the WF are (see Appendix): 
\begin{equation}\label{eq:Wigner_bessel_zero_order}
W(R) 
\, \approx \, 
\frac{e^{-R^2-d^2}}{\pi N} \left[ (1-\Delta)\, e^{-d^2} I_0(r) + \Delta \, M \, J_{0}(r) \right],
\end{equation}
where $J_{0}(r)$ and $I_0(r)$ are the standard and modified Bessel functions of the first kind, respectively~\cite{abramowitz+stegun}.
Interestingly, in the limit of highly packed states, the WF loses the angular dependency and depends only on the radial variable, while the approximation $R {<} M/(2\sqrt{2}d)$ allows us to express the off-diagonal terms as a single Bessel function. In the absence of dephasing, i.e., $\Delta = 1$, the behavior of the WF simplifies and turns out to be determined only by $J_{0}(r)$. This result is confirmed by the numerical simulation shown in Fig.~\ref{fig:fig2} a), where the agreement between numerical results and analytical predictions remains robust for $M \ge d^2$.

Solving $(1-\Delta)\, e^{-d^2} I_0(r) + \Delta \, M \, J_{0}(r)=0$ for $\Delta$ and minimizing over $r$ provides us the upper bound $\overline{\Delta}_c$ of $\Delta_c$ for $d \ge 1$:
\begin{equation}\label{eq:delta_c_high_order}
    \overline{\Delta}_c = \min_{r}\frac{1}{1- e^{d^2}  M \frac{J_0(r)}{I_0(r)}} = \frac{1}{1 - e^{d^2}  M\frac{J_0(\pi)}{I_0(\pi)}} = \frac{1}{1+0.056 \,e^{d^2}  M},
\end{equation}
where the minimum is for $r \approx 3.2$ at which $J_0(r)/I_0(r) \approx -0.056$ [see Fig.~\ref{fig:fig2} a) and see Appendix for more details]. Therefore, if $\Delta \ge \overline{\Delta}_c$, the Wigner function is negative.

In the highly packed limit, the upper bound $\overline{\Delta}_c$ well-approximates $\Delta_c$ (boundary between positive and negative regions of the WF), as it exhibits the same trend with $d$. This is shown in Fig.~\ref{fig:fig2} b), where we compare $\overline{\Delta}_c$ with the values of $\Delta_c$ computed numerically by searching for the minimum of WF over the full $X$-$P$ space.

Similarly to the previous case, when $d$ is much larger compared to the characteristic length scale of the problem ($d \gg \sigma$), $\Delta_c$ becomes exponentially small. As a result, in such a regime, the presence of coherent superpositions in the system's density matrix within the CSB practically represents a necessary and sufficient condition for the WF negativity.

{\it Conclusions---}
We have introduced an operational criterion to determine the negativity of the WF for a generic density matrix. This criterion is associated with a well-defined protocol (the QNDM scheme) that has already found application in many contexts~\cite{GuerlinNature2007,Deleglise2008,GeremiaPRL2003,GuerlinNature2007,solinas2021,NiemietzNature2021,solinas2022,YanagimotoPRXQuantum2023} and it identifies the coherent states basis as a privileged one to establish the condition for a negative WF. Starting from Refs.~\cite{solinas2024,SolinasReview2026}, we studied to what extent the presence of coherent superpositions in this basis provides direct information about the WF negativity.

Sufficient condition for the WF negativity can be readily proven; that is, the absence of coherent-basis superpositions implies the positivity of the WF. Then, for the Schr\"{o}dinger-cat states and high-order cat states on a circle, relevant case-studies in the literature, we also discuss the necessary condition that guarantees negativity. In fact, we identify a critical residual coherence above which the WF exhibits negativity. This result is analytically exact for the Schr\"{o}dinger-cat state, and approximate for high-order cat states.

The presence of a critical residual coherence $\Delta_c$ is directly related to the decomposition of the quantum state in the CSB, which is continuous and overcomplete. However, given the exponential reduction of $\Delta_c$ with the system’s length scale, in many cases one can argue that the mere presence of quantum superpositions in the CSB yields a negative WF.

Our results provide a new perspective, both theoretically and experimentally, for analyzing the WF and its properties. These ideas could be exploited to track the quantum-to-classical transition in many experimental set-ups~\cite{Arndt1999, GuerlinNature2007, Deleglise2008, Gerlich2011, Vlastakis2013, Aspelmeyer2022, FuchsScienceAdvances2024, Pedalino2026}, or used to certify the presence of quantum effects in quantum continuous variable systems~\cite{Mu1996, Braunstein2005, Weedbrook2012, Adesso2014}. Nevertheless, a significant question remains open, namely, to understand whether the necessary condition can be extended to more complex discrete configurations or to the continuous case, while still dealing with an arbitrary density matrix. 

\begin{acknowledgments}
{\it Acknowledgements---} 
B.D.~and S.G.~acknowledge financial support from the PNRR MUR project PE0000023-NQSTI funded by the European Union---Next Generation EU. 

{\it Authors' contributions---} 
P.S.~developed the idea and carried out the first set of calculations. B.D.~performed the numerical simulations. All authors contributed to the analytical calculations, discussed the results, and wrote the manuscript.
\end{acknowledgments}

\pagebreak
\widetext
\begin{center}
\textbf{\large Appendix}
\end{center}
\setcounter{equation}{0}
\setcounter{figure}{0}
\setcounter{table}{0}
\setcounter{page}{1}
\makeatletter
\renewcommand{\theequation}{S\arabic{equation}}
\renewcommand{\thefigure}{S\arabic{figure}}

\section{Quasi-characteristic function, and quantum non-demolition measurements}
\label{app:WF_QNDM}

As introduced in the main text, consider a tripartite system composed of a quantum system $\mathcal{S}$ and two detectors $\mathcal{D}_1$ and $\mathcal{D}2$, with operators $\hat\lambda$ and $\hat\eta$ acting on them, respectively. The operators of position $\hat x$ and momentum $\hat p$ act on $\mathcal{S}$.

We take $\hat\rho$ as the initial state of $\mathcal{S}$, and $\hat\rho_1 = \frac{1}{\sqrt{M_1}} \sum_{\lambda,\lambda'} \ketbra{\lambda}{\lambda'}$ for $\mathcal{D}1$ and $\hat\rho_2 = \frac{1}{\sqrt{M_2}} \sum_{\eta,\eta'} \ketbra{\eta}{\eta'}$ for $\mathcal{D}_2$, with $M_1$ and $M_2$ appropriate normalization constants. The total initial density matrix is $\hat R_0 = \hat\rho \otimes \hat\rho_1 \otimes \hat\rho_2$. Denoting by $\mathrm{Tr}_{\mathcal{S}}$ the trace over the system’s degrees of freedom, the off-diagonal elements of $\hat R_0$ between the two detectors' states $\ket{\frac{\lambda}{2}, \frac{\eta}{2}}$ and $\ket{-\frac{\lambda}{2}, -\frac{\eta}{2}}$ is
\begin{equation}
     \Trace_{\mathcal{S}} \left[ \left\langle\frac{\lambda}{2}, \frac{\eta}{2}\right| \hat{R}_0 \left| -\frac{\lambda}{2}, -\frac{\eta}{2}\right\rangle\right] 
    = \frac{1}{M_1 M_2}.
\end{equation}

The system-detector coupling is generated by the unitary operator $\hU_{\lambda,\eta}  = \exp{ i (\hx \otimes \hlambda + \hp \otimes \heta) }$, where $\hx \otimes \hlambda \equiv \hx \otimes \hlambda \otimes \Idoperator$ and $\hp \otimes \heta \equiv \hp \otimes \Idoperator \otimes \heta$. The density matrix of the tripartite system after the coupling is $\hat{R}_f=  \hU_{\lambda,\eta} \hat{R}_0 \hU^\dagger_{\lambda,\eta}$. The phase accumulated between $\ket{\frac{\lambda}{2}, \frac{\eta}{2}}$ and $\ket{-\frac{\lambda}{2}, -\frac{\eta}{2}}$ during the evolution is defined as
\begin{equation}
    \G_{\lambda,\eta} = \frac{\Trace_{\mathcal{S}} \left[ \left\langle\frac{\lambda}{2}, \frac{\eta}{2}\right| \hat{R}_f \left| -\frac{\lambda}{2}, -\frac{\eta}{2} \right\rangle\right]}{\Trace_{\mathcal{S}} \left[ \left\langle\frac{\lambda}{2}, \frac{\eta}{2}\right| \hat{R}_0 \left| -\frac{\lambda}{2}, -\frac{\eta}{2}\right\rangle\right]}\,.
    \label{app_eq:accumulated_phase}
\end{equation}
The numerator of \eqref{app_eq:accumulated_phase} reads as
\begin{eqnarray}
    &&\Trace_{\mathcal{S}} \left[ \left\langle\frac{\lambda}{2}, \frac{\eta}{2}\right| \hat{R}_f \left| -\frac{\lambda}{2}, -\frac{\eta}{2}\right\rangle \right] =
    \Trace_{\mathcal{S}} \left[ \left\langle\frac{\lambda}{2}, \frac{\eta}{2}\right| \hU_{\lambda,\eta} \hat{R}_0 \hU^\dagger_{\lambda,\eta} \left| -\frac{\lambda}{2}, -\frac{\eta}{2}\right\rangle \right]=\nonumber\\
    &&=
    \Trace_{\mathcal{S}} 
    \left[
    e^{ \frac{i}{2} (\lambda \hx + \eta   \hp) }
    \left\langle \frac{\lambda}{2}, \frac{\eta}{2} \right| \hat{R}_0 \left| -\frac{\lambda}{2}, -\frac{\eta}{2}\right\rangle
    e^{ \frac{i}{2} (\lambda \hx + \eta \hp) }
    \right] = \frac{1}{M_1 M_2}\Trace_{\mathcal{S}} \left[
    e^{ i (\lambda \hx+ \eta   \hp) }
    \hrho
    \right].
\end{eqnarray}
As a result, the phase accumulated in \eqref{app_eq:accumulated_phase} is the quasi-characteristic function of the WF:
\begin{equation}
    \G_{\lambda,\eta} = \Trace_{\mathcal{S}} \left[ e^{ i (\lambda \hx+ \eta   \hp) }
    \hrho \right].
\end{equation}

By using the Zassenhaus formula and the standard commutation relation $[\hx , \hp] =i $, we have:
\begin{equation}
\exp\left\{ i (\lambda \hx + \eta   \hp) \right\} =
\exp\left\{ i \lambda \eta/2 \right\} \exp\left\{ i \lambda \hx \right\} \exp\left\{ i  \eta   \hp \right\} = \exp\left\{ -i \lambda \eta/2 \right\} \exp\left\{ i  \eta   \hp \right\} \exp\left\{ i \lambda \hx \right\}.
\end{equation}
Hence, using the properties of the trace,  
\begin{equation}
    \G_{\lambda,\eta} = \Trace_{\mathcal{S}} \left[
    e^{ i (\frac{\lambda}{2} \hx + \frac{\eta}{2}   \hp) }
    \hrho \, e^{ i (\frac{\lambda}{2} \hx + \frac{\eta}{2}   \hp) } 
    \right] 
    = \Tr_{\mathcal{S}}  \left[ e^{ \frac{i\lambda}{2} \hx} e^{\frac{i\eta}{2}\hp } \hrho \,  e^{\frac{i\eta}{2}\hp } e^{\frac{i\lambda}{2} \hx} \right],
    \label{app_eq:G_expanded}
\end{equation}
which corresponds to \eqref{eq:G_def} in the main text. It is worth stressing that the usual formulation for $\G_{\lambda,\eta}$ as a function of parameters $\lambda$ and $\eta$ is~\cite{solinas2015fulldistribution}:
\begin{equation}
\G_{\lambda,\eta} = \Trace \left[ \hat{\mathcal{U}}_{\lambda/2} \hat{\mathcal{U}}_{\eta/2} \, \hrho \, \hat{\mathcal{U}}^\dagger_{-\eta/2} \hat{\mathcal{U}}^\dagger_{-\lambda/2} \right].
\end{equation}
Such expression is exactly in the form of \eqref{app_eq:G_expanded} with $\hat{\mathcal{U}}_{\eta/2} = \exp\left\{ \frac{i\eta}{2}\hp \right\}$ and $\hat{\mathcal{U}}_{\lambda/2} = \exp\left\{ \frac{i\lambda}{2} \hx \right\}$. In the framework of quantum non-demolition measurements, \eqref{app_eq:G_expanded} can be interpreted as obtained from a sequential measurement of the momentum and position operators~\cite{solinas2015fulldistribution, solinas2016probing,solinas2024,GherardiniTutorial,SolinasReview2026}.

\section{Invariance of the Wigner function under length rescaling}
\label{app:WF_invariance}

Let us consider the definition of the Wigner Function (WF) in terms of the quasi-characteristic function:
\begin{equation}\label{eq:WF_def_CharFun}
W(X,P) = \frac{1}{4\pi^2} \int d\lambda~d\eta \, e^{ -i \lambda X} e^{ -i\eta P} \Tr[e^{ i (\lambda \hx+ \eta   \hp) } \hrho]. 
\end{equation}
Following Ref.~\cite{Diosi2002}, we rescale all the lengths by a factor $c$: $X \rightarrow c X$, $P \rightarrow P/c$, $\hx \rightarrow c \hx$, $\hp \rightarrow \hp/c$, such that the rescaled Wigner function reads as
\begin{equation}
W \left(c X,\frac{P}{c} \right) =  
\frac{1}{4 \pi^2} \int d\lambda~d\eta~ e^{ -i c \lambda X}  e^{ -i\eta \frac{P}{c}} \Tr[e^{ i (c \lambda \hx + \eta \frac{\hp}{c}) } \hrho]. 
\end{equation}
By changing the integration variables to $\Lambda = c \lambda$ and $\Gamma = \eta/c$, the extremes of integration do not change, and thus we can write:
\begin{equation}
W \left(c X,\frac{P}{c} \right) =  
\frac{1}{4 \pi^2} \int d\Lambda~d\Gamma~ e^{ -i  \Lambda X}  e^{ -i\Gamma P} \Tr[e^{ i ( \Lambda \hx+ \Gamma \hp) } \hrho] = W(X,P). 
\end{equation}
Therefore, the Wigner function is invariant under canonical linear transformation. This result was already discussed by Di\'osi and Kiefer in Ref.~\cite{Diosi2002}.

\section{Sufficient condition for the Wigner function negativity}
\label{app:WF_positivity}

Let us introduce the creation and annihilation operators that are linear combinations of $\hx$ and $\hp$:
\begin{eqnarray}
	\has &=& \frac{1}{\sqrt{2}\sigma} \Big( \hx +  i \sigma^2 \hp \Big) \label{app_eq:a} \\  
	\hasdg &=& \frac{1}{\sqrt{2}\sigma} \Big( \hx - i \sigma^2 \hp \Big), \label{app_eq:adg}	
\end{eqnarray}
where $\hx = \frac{\sigma}{\sqrt{2} } \left(\hasdg + \has \right)$ and $\hp = \frac{i  }{\sqrt{2} \sigma} \left(\hasdg - \has\right )$ are the inverse relations. Eqs.~(\ref{app_eq:a})-(\ref{app_eq:adg}) obey the usual commutation relations $[\has, \hasdg] = 1$, given that $[\hx, \hp] = i$ ($\hbar=1$).

In the following, we rescale all the lengths over $\sigma$ that is equivalent to take $\sigma=1$ in the above formulas. The exponential operator $e^{i( \eta \hp+ \lambda \hx)}$ can be written as 
\begin{eqnarray}
	e^{i( \eta \hp + \lambda \hx)} 
    &=& \exp \left\{ \frac{i \lambda - \eta }{\sqrt{2}} \hadg +
	\frac{i \lambda  +   \eta }{\sqrt{2} } \ha \right\} = \exp \left\{ \frac{i \lambda - \eta }{ \sqrt{2} } \hadg \right\}
	\exp \left\{ \frac{i \lambda + \eta }{ \sqrt{2} } \ha \right\}
	\exp \left\{ -\frac{\lambda^2 + \eta^2 }{4} \right\} =\nonumber \\
    &=&
    \exp \left\{ \alpha \hadg \right\}
	\exp \left\{ -\alpha^* \ha \right\}
	\exp \left\{ -\frac{|\alpha|^2}{4} \right\},
\end{eqnarray}
where $\alpha \equiv (i \lambda -   \eta)/\sqrt{2}$, $\re \alpha = - \eta/\sqrt{2}$ and $\im \alpha = \lambda/\sqrt{2}$.

If the system's density matrix is diagonal in the coherent state $\ket{\beta}$, we can write it as $\hrho = \int d^2 \beta ~ \rho_{\beta \beta} \ketbra{\beta}{\beta}$ with $d^2 \beta = d \re \beta~d \im \beta$ and $\rho_{\beta \beta}=\braket{\beta|\hrho|\beta}\geq 0$~\cite{Glauber1963}. Therefore, the quasi-characteristic function simplifies to
\begin{eqnarray}
	\G_{\lambda, \eta} = e^{-\frac{|\alpha|^2}{2}} \Trace \left[ e^{-\alpha^* \ha} \hrho \, e^{\alpha \hadg} \right] = e^{-\frac{|\alpha|^2}{2}} \int d^2 \beta~\rho_{\beta \beta} \, e^{-\alpha^* \beta + \alpha \beta^*},
\end{eqnarray}
where we have used the fact that the coherent states are eigenstates of $\ha$, i.e., $e^{-\alpha^* \ha} \ket{\beta}= e^{-\alpha^* \beta} \ket{\beta}$ and $\bra{\beta}e^{\alpha \hadg} = \bra{\beta} e^{\alpha \beta^*}$, and that $\Trace \left[ \ketbra{\beta}{\beta} \right] = 1$.
Expressing the integrals in $\lambda$ and $\eta$ in terms of $\re\alpha$ and $\im\alpha$, the WF $W(X,P)=\frac{1}{4\pi^2}\int d\lambda~d\eta \, e^{ -i \lambda X} e^{ -i\eta P}\G_{\lambda, \eta}$ can be written as 
\begin{equation}
    W(X, P) = \frac{1}{2 \pi^2} \int d^2 \beta~\rho_{\beta \beta}
    \int d\re\alpha~  
    e^{-\frac{(\re\alpha)^2}{2}}
    e^{i \sqrt{2} \, \re\alpha \left[P - \sqrt{2} \, \im\beta \right]} 
    \int d\im\alpha~
    e^{-\frac{(\im\alpha)^2}{2}}
    e^{-i \sqrt{2} \, \im\alpha \left[X - \sqrt{2} \, \re\beta \right]}.
\end{equation}
Thus, by performing the Gaussian integrations, we obtain: 
\begin{equation}\label{eq:positive_WF}
    W(X, P) = \frac{1}{\pi}\int d^2 \beta~\rho_{\beta\beta} \, e^{-\left(X - \sqrt{2}\,\re\beta \right)^2} e^{- \left(P - \sqrt{2}\,\im\beta \right)^2}.
\end{equation}
Since all the integrand terms are positive, the WF (\ref{eq:positive_WF}) is positive.

\subsection{Remark on the sufficient condition for Wigner function negativity}

Albeit straightforward, we would like to stress that the result about the presence of a privileged basis, i.e., the coherent state basis, and the sufficient condition for WF negativity is not as trivial as it might seem. Even for one-dimensional systems, indeed, there are three natural bases in which to decompose the system's density matrix associated with the operators $\hat x$, $\hat p$ and $\hat a$, the latter being a linear combination of $\hat x$ and $\hat p$. Any diagonal density matrix in one of these bases would lead to a positive WF. However, a density matrix diagonal in the $\hat x$ basis would yield an unphysical and unnormalized WF.

To show this, we start from an alternative definition of the WF~\cite{WignerPhysRev1932, Hillery1984}: $W(X,P) = \int dY e^{i Y P} \braket{X-\frac{Y}{2} | \hrho | X + \frac{Y}{2} }$, and we assume that the density matrix is diagonal in the position basis, i.e., $\hat\rho = \int dx \rho(x,x)\ketbra{x}{x}$. Using the orthogonality of position eigenstates, i.e., $\braket{x|y} = \delta(x-y)$, we obtain:
\begin{equation}\label{app_eq:WF_nonphysical}
    W(X,P) = \int dY e^{i Y P}\int  dx \,  \rho(x,x) \, \delta\left(x-X+\frac{Y}{2}\right) \delta\left(x-X-\frac{Y}{2}\right) = \rho(X,X).    
\end{equation}
The WF (\ref{app_eq:WF_nonphysical}) is nonphysical because it is not normalizable: the integration over $P$ is not bounded. Indeed, it is known in the literature that the WF cannot be arbitrarily squeezed in phase space~\cite{Hillery1984}, as it is in \eqref{app_eq:WF_nonphysical}.

In conclusion, despite formally leading to positive WF, the cases in which the density matrix becomes diagonal in the basis of $\hat x$ or $\hat p$ are unphysical and must be regarded as mathematical limits. This observation is related to the findings in Refs.~\cite{Diosi2000, Diosi2002, Brody2025}, where, for open systems, it is shown that the density matrix reaches a diagonal form only in the mathematical limit of infinite-time evolution.

\section{Wigner function for a superposition of coherent states}

Let us start again from the definition of the WF in terms of the quasi-characteristic function [see \eqref{eq:WF_def_CharFun}] and substitute a generic expression for the density matrix $\hat{\rho} = (1/N)\sum_{i,j} \rho_{ij} \ketbra{i}{j}$. We obtain:
\begin{align}
    W(X,P) &= \frac{1}{4 \pi^2} \int d\lambda~d\eta~ e^{ -i \lambda X}  e^{ -i\eta P} \, \text{Tr} \left[ e^{ i (\lambda \hx+ \eta   \hp) } \frac{1}{N} \sum_{i,j} \rho_{ij} \ketbra{i}{j} \right] = \\
    &= \frac{1}{N} \sum_{i,j} \rho_{ij}  \left\{ \frac{1}{4 \pi^2} \int d\lambda~d\eta~ e^{ -i \lambda X}  e^{ -i\eta P} \, \text{Tr} \left[ e^{ i (\lambda \hx+ \eta   \hp) }\ketbra{i}{j} \right] \right\} = \frac{1}{N} \sum_{i,j} \rho_{ij} W_{ij} \,,
\end{align}
where $W_{ij}$ is the cross-Wigner function associated with the states $\ket{i}$ and $\ket{j}$. For $i=j$, the cross-Wigner function reduces to the standard Wigner function of the state $\ket{i}$. For $i\neq j$, these terms correspond to interference contributions in phase space. We can rewrite it by isolating the cross-Wigner from the standard ones
\begin{equation}\label{eq:Wigner_sum}
    W(X,P) = \frac{1}{N} \sum_{i} \rho_{ii} W_{ii}(X,P) + \frac{1}{N} \sum_{i \neq j} \rho_{ij} W_{ij}(X,P) \equiv W_{D}(X,P) + W_{OD}(X,P) \,,
\end{equation}
where we have defined the diagonal and off-diagonal elements of the total Wigner function. 

\section{Schr\"{o}dinger-cat states: Necessary condition for Wigner function negativity}\label{app:cat_states}

We denote the coherent state associated with the complex number $\beta$ and variance $\sigma^2$ as $\ket{\beta, \sigma}$. With this notation, let us consider a system prepared in a superposition of two coherent states.
The density matrix can be written as 
\begin{eqnarray}
	\hat{\rho} = \frac{1}{N} \Big( \rho_{1,1} \ketbra{\beta_1, \sigma}{\beta_1, \sigma} + \rho_{2,2} \ketbra{\beta_2, \sigma}{\beta_2, \sigma}+
	\rho_{1,2} \ketbra{\beta_1, \sigma}{\beta_2, \sigma} + \rho_{2,1} \ketbra{\beta_2, \sigma}{\beta_1, \sigma} \Big),
\end{eqnarray}
where the normalization constant is
\begin{equation}
	N = \rho_{1,1} +  \rho_{2,2} + \rho_{1,2} e^{-\frac{|\beta_1|^2 + |\beta_2|^2}{2}}  e^{\beta_1 \beta_2^*} + \rho_{2,1} e^{-\frac{|\beta_1|^2 + |\beta_2|^2}{2}}  e^{\beta_2 \beta_1^*} \,.
\end{equation}
As discussed in the main text, we restrict our analysis to the case $X_{1,2} = \pm \sqrt{2}\,\text{Re}\beta$ and $P_{1,2}=0$. Using the expression of the Wigner function as a sum [see \eqref{eq:Wigner_sum}], we obtain two diagonal contributions ($i=j=1,2$):
\begin{equation}
    W_{1,1} = \frac{1}{\pi} e^{-P^2} e^{-(X-\sqrt{2} \text{Re}\beta)^2}; \qquad
    W_{2,2} = \frac{1}{\pi} e^{-P^2} e^{-(X+\sqrt{2} \text{Re}\beta)^2},
\end{equation}
and two off-diagonal contributions ($i=1,j=2$ and $i=2,j=1$):
\begin{equation}
     W_{1,2} = \frac{1}{\pi} e^{-P^2} e^{-X^2} e^{-i 2\sqrt{2} \text{Re}\beta P} = W_{2,1}^* \,.
\end{equation}
Using the self-adjoint property of the off-diagonal elements, the total Wigner function can be written as
\begin{align}
    W &= \frac{1}{N} \left[\rho_{1,1}W_{1,1} + \rho_{2,2}W_{2,2} + \rho_{1,2}W_{1,2} + \rho_{2,1}W_{2,1} \right] = \nonumber \\
    &= \frac{1}{N} \left[ \rho_{1,1}W_{1,1} + \rho_{2,2}W_{2,2} + 2 \re (\rho_{1,2}W_{1,2}) \right].
\end{align}
Let us now parametrize the density matrix with the angles $\theta$ and $\Delta$ as
\begin{equation}
    \hat{\rho} = \frac{1}{N}
    \begin{pmatrix}
        \cos^2{\theta} & \Delta\sin{\theta}\cos{\theta}e^{i\phi} \\
        \Delta\sin{\theta}\cos{\theta}e^{-i\phi} & \sin^2{\theta}
    \end{pmatrix}
\end{equation}
where $0 \leq \Delta \leq 1$ is a parameter that accounts for the degree of coherence of the system.

Substituting it into the Wigner function yields:
\begin{align}
    W &= \frac{1}{\pi N} \left[ \cos^2{\theta} \, e^{-P^2} e^{-(X-\sqrt{2} \text{Re}\beta)^2} + \sin^2{\theta} \, e^{-P^2} e^{-(X+\sqrt{2} \text{Re}\beta)^2} 
    + 2 \Delta \re\left( \sin{\theta}\cos{\theta}e^{i\phi} e^{-P^2} e^{-X^2} e^{-i 2\sqrt{2} \text{Re}\beta P} \right) \right] = \nonumber \\
    &= \frac{e^{-X^2-P^2}}{\pi N} \left[ \cos^2{\theta} \, e^{2\sqrt{2} \text{Re}\beta X -2\text{Re}\beta^2 } + \sin^2{\theta} \, e^{-2\sqrt{2} \text{Re}\beta X -2\text{Re}\beta^2}
    + \Delta |\sin{2\theta}| \cos{\left(\phi - 2\sqrt{2} \text{Re}\beta P \right)} \right] = \nonumber \\
    &= \frac{e^{-X^2-P^2-2\text{Re}\beta^2}}{\pi N} \left[ \cos^2{\theta} \, e^{2\sqrt{2} \text{Re}\beta X } + \sin^2{\theta} \, e^{-2\sqrt{2} \text{Re}\beta X }
    + e^{2\text{Re}\beta^2} \Delta |\sin{2\theta}| \cos{\left(\phi - 2\sqrt{2} \text{Re}\beta P \right)} \right] \,.
\end{align}
Defining $\tphi\equiv\phi-2\sqrt{2}\text{Re}\beta P$ and $Z\equiv e^{2\sqrt{2}\text{Re}\beta X}$, we obtain that
\begin{align}
    W &= \frac{e^{-X^2-P^2-2\text{Re}\beta^2}}{\pi N} \left[ \cos^2{\theta} \, Z + \sin^2{\theta} \, Z^{-1} + e^{2\text{Re}\beta^2} \Delta |\sin{2\theta}| \cos{\tphi} \right] = \\
    &= \frac{e^{-X^2-P^2-2\text{Re}\beta^2}}{\pi N Z} \left[ \cos^2{\theta} \, Z^2 + b \, Z + \sin^2{\theta} \right] \,,
\end{align}
where we have introduced $b\equiv e^{2(\text{Re}\beta)^2} \Delta |\sin{2\theta}| \cos{\tphi}$. 
The last expression coincides with Eq.\,(5)
of the main text. The zeros of the Wigner function are obtained by imposing that the quadratic polynomial inside the brackets vanishes, i.e. $\mathcal{P}(Z)=\cos^2{\theta} \, Z^2 + b \, Z + \sin^2{\theta}=0$, yielding
\begin{equation}
    Z_{\pm} = \frac{-b\pm \sqrt{b^2-4\sin^2{\theta} \cos^2{\theta}}}{2\cos^2{\theta}} \,.
\end{equation}
Since $\cos^2\theta >0$, the parabola $\mathcal{P}(Z)$ is concave up, and its minimum occurs at the middle point of the two solutions $Z_m = (Z_+ + Z_-)/2$. Therefore, one finds that the parabola at its vertex assumes the minimum value
\begin{equation}
    \mathcal{P}(Z_m) = 2 \sin^2\theta \left(1- \Delta^2 \cos^2 \tphi~e^{4 \re \beta^2} \right) \,.
\end{equation}
As the prefactor $(e^{-X^2-P^2 - 2 \re \beta^2})/(\pi N Z)$ is non-negative, the condition to have a negative WF for a cat state is given by the inequality $\mathcal{P}(Z_m)<0$, which corresponds to $\Delta > e^{-2 \re \beta^2}/|\cos\tphi|$. We have excluded the case $\sin{\theta}=0$, which would correspond to a density matrix with all the population in the first coherent state and zero population in the other one. 

The strongest bound is obtained when the cosine term is maximal, i.e., $\cos\tphi=1$, corresponding to momenta satisfying $\tphi=2\pi k$. In this case, the condition for Wigner-function negativity reduces to
\begin{equation}
    \Delta > \Delta_c \equiv e^{-2(\mathrm{Re}\,\beta)^2},
\end{equation}
which coincides with \eqref{eq:Delta_c} of the main text. 

\section{Higher-order cat states on a circle}
\label{app:multiple_cat_states}

Let us consider the Wigner function of $M$ (even) equally weighted coherent states arranged on a circle of radius $d$. The specific geometry of the state makes it natural to choose to work in polar coordinates. The positions of the centers of the $M$ coherent states in phase-space are defined as points on a circle of radius $d$ with equally spaced angular position, i.e. $\{ X_k, P_k \} = \sqrt{2} d \{ \cos \theta_k, \sin \theta_k\}$ with $\theta_k = 2 \pi k/M$ and $1 \leq k \leq M$, while the polar coordinates are defined as $\{X, P \} = \{R \cos \phi, R \sin \phi \}$. 
We are interested in the highly packed states limit, which formally corresponds to $M \gg d^2$. Similarly to the cat-state case, we parametrize the density matrix with a coherence factor $\Delta$ multiplying the off-diagonal elements $\rho_{j,k} = [(1-\Delta) \delta_{jk} +\Delta]/M$ with $0 \leq \Delta \leq 1$ equal for all the off-diagonal elements.

Let us restart from the total Wigner function, defined in Eqs.~(3)-(4) of the main text, and rewrite it in polar coordinates using the above definitions
\begin{align}
    W(R,\phi) &= \frac{e^{-R^2-d^2}}{\pi N M} \sum_{j=1}^{M}\sum_{k=1}^{M} \left[ (1-\Delta)\delta_{jk} + \Delta \right] \exp\left( -d^2 e^{2 i \Phi_{j,k}} + r \cos(\phi-\Lambda_{j,k}) e^{i\Phi_{j,k}} \right)=\nonumber \\
    &=(1-\Delta)\frac{e^{-R^2-2d^2}}{\pi N M} \sum_{j=1}^{M} e^{r \cos(\phi-\theta_j)} + 
    \Delta\frac{e^{-R^2-d^2}}{\pi N M} \sum_{j,k=1}^{M} e^{-d^2 e^{i2\Phi_{j,k}}+r \cos(\phi-\Lambda_{j,k}) e^{i\Phi_{jk}}} \,,
\end{align}
where we have defined $r=2\sqrt{2}Rd$, $\Lambda_{j,k}=(\theta_j+\theta_k)/2$ and $\Phi_{j,k}=(\theta_j-\theta_k)/2$. 
Introducing the index $\ell\equiv j-k$, we obtain $\Phi_\ell=\pi \ell/M$ and $\Lambda_{\ell,k}=(\ell +2k)\pi/M$, so that
\begin{equation}
    W(R, \phi) = (1-\Delta)\,  \frac{e^{-R^2 - 2d^2}}{\pi N M} \sum_{j=1}^{M} e^{r \cos(\phi - \theta_j)} + \Delta \, \frac{e^{-R^2 - d^2}}{\pi N M} \sum_{k=1}^{M} \sum_{\ell=0}^{M-1} e^{-d^2 e^{2i \Phi_\ell}} \, e^{r e^{i \Phi_\ell} \cos(\Lambda_{\ell,k} - \phi)}
\end{equation}
It is worth noting that this parametrization is valid only if the function $f(j,k)$, defined as the argument of the double sum, is periodic with period $M$ in its first index. In particular, we require $f(j+M,k)=f(j,k)$, indeed
\begin{align}
    f(j+M,k) &= \exp{{-d^2 e^{i\frac{2\pi}{M}(j-k+M)}} \, + r e^{i \frac{\pi}{M}(j-k+M)} \cos \left(\phi - \frac{\pi}{M} (j+k+M) \right) } = \nonumber \\
    &= \exp{{-d^2 e^{i\frac{2\pi}{M}(j-k)}} \, + r (-1) e^{i \frac{\pi}{M}(j-k)} \cos \left(\phi - \frac{\pi}{M} (j+k) -\pi \right) } = f(j,k) \,,
\end{align}
where we have used $e^{i\pi}=-1$ and $\cos{(x-\pi)=-\cos(x)}$. This ensures that the two sums can be treated as independent. 
At this point, we perform a Jacobi-Anger expansion  over the modified Bessel functions \cite{abramowitz+stegun}: 
\begin{equation}
    e^{\alpha_\ell \cos(\Lambda_{\ell,k} - \phi)} = 
    \sum_{n=-\infty}^{+\infty} I_n(\alpha_\ell) \, e^{i n (\Lambda_{\ell,k} - \phi)} 
\end{equation}
where $\alpha_\ell\equiv r e^{i \Phi_\ell}$. 
Using the identity
\begin{equation}\label{eq:geometric_serie}
    \sum_{p=1}^{M} \left(e^{i n \frac{2\pi}{M}}\right)^p = 
    \begin{cases}
    M & \text{if } n = qM \quad \text{with } q \in \mathbb{Z} \\
    0 & \text{else}
    \end{cases} \,,
\end{equation}
we find that the sum over $k$ becomes 
\begin{equation}
    \sum_{k=1}^{M} e^{i n \Lambda_{\ell,k} } = e^{in\ell \pi/M } \sum_{k=1}^{M} e^{i n k 2\pi/M} = e^{in\ell \pi/M } M \delta_{n,qM} \,.
\end{equation}
Hence, the double sum can be rewritten as
\begin{eqnarray}
    && \sum_{k=1}^{M} \sum_{\ell=0}^{M-1} e^{-d^2 e^{2i \Phi_\ell}} \, e^{r e^{i \Phi_\ell} \cos(\Lambda_{\ell,k} - \phi)} = \sum_{k=1}^{M} \sum_{\ell=0}^{M-1} e^{-d^2 e^{2i \Phi_\ell}} \,  \sum_{n=-\infty}^{+\infty} I_n(\alpha_\ell) \, e^{i n (\Lambda_{\ell,k} - \phi)} = \nonumber \\
    &&= \sum_{\ell=0}^{M-1} e^{-d^2 e^{2i \Phi_\ell}} \,  \sum_{n=-\infty}^{+\infty} I_n(\alpha_\ell) \, e^{-i n \phi} e^{in\ell \pi/M } M \delta_{n,qM} = \sum_{\ell=0}^{M-1} e^{-d^2 e^{2i \Phi_\ell}} \,  \sum_{q=-\infty}^{+\infty} I_{qM}(\alpha_\ell) \, e^{-i qM \phi} e^{iq\ell \pi} M = \nonumber \\
    &&= \sum_{\ell=0}^{M-1} e^{-d^2 e^{2i \Phi_\ell}} \,  \sum_{q=-\infty}^{+\infty} I_{qM}(\alpha_\ell) \, e^{-i qM \left( \phi - \frac{\pi}{M} \ell \right)} M.
\end{eqnarray}
An analogue calculation applies also to the diagonal contribution
\begin{equation}
    \sum_{j=1}^{M} e^{r \cos(\phi - \theta_j)} = \sum_{j=1}^{M} \sum_{n=-\infty}^{+\infty} I_{n}(r) \, e^{i n (\phi - \theta_j)} = \sum_{n=-\infty}^{+\infty} I_{n}(r) \, e^{in\phi} \sum_{j=1}^{M} e^{i n \frac{2\pi j}{M}} = \sum_{n=-\infty}^{+\infty} I_{n}(r) \, e^{in\phi} M \delta_{n,qM} = \sum_{q=-\infty}^{+\infty} I_{qM}(r) \, e^{i qM \phi} M.
\end{equation}
Putting all the calculations together, the Wigner function becomes:
\begin{equation}
    W(R, \phi) = (1-\Delta)\,  \frac{e^{-R^2 - 2d^2}}{\pi N } 
    \sum_{q=-\infty}^{+\infty} I_{qM}(r) \, e^{i qM \phi} + \Delta \, \frac{e^{-R^2 - d^2}}{\pi N} \sum_{\ell=0}^{M-1} e^{-d^2 e^{2i \frac{\pi}{M} \ell}} \, \sum_{q=-\infty}^{+\infty} I_{qM}(r e^{i \frac{\pi}{M} \ell}) \, e^{-i qM \left( \phi -\frac{\pi}{M} \ell \right)}.
\end{equation}
In both diagonal and off-diagonal contributions, the angular dependence appears through terms of the form
\begin{equation}
    \sum_{q=-\infty}^{+\infty} I_{qM}( x) \, e^{i qM y} \,,
\end{equation}
which reflects the underlying rotational symmetry of the state.

We now identify the dominant contribution to the sum over modified Bessel functions. As known, in the limit $x\rightarrow0$ the only surviving contribution is the zero Bessel function $I_0(x)$. More precisely, for $0<\abs{\alpha_{\ell}} \ll \sqrt{qM+1}$, we have
\begin{equation}
    I_{qM}(\alpha_\ell) \sim \frac{1}{\Gamma(qM+1)} \left( \frac{\alpha_\ell}{2} \right)^{qM} = \frac{1}{(qM)!} \left( \frac{\alpha_\ell}{2} \right)^{qM}.
\end{equation}
Using the Stirling's approximation $n! \sim \sqrt{2\pi n}(n/e)^n$, $I_{qM}(\alpha_\ell)$ simplifies to
\begin{equation}
     I_{qM}(\alpha_\ell) \sim \left( \frac{\alpha_\ell}{2} \right)^{qM} \frac{1}{\sqrt{2\pi qM}} \left( \frac{e}{qM} \right)^{qM} 
     = \frac{1}{\sqrt{2\pi qM}} \left( \frac{e \, \alpha_\ell}{2 qM} \right)^{qM}
     \underset{qM\rightarrow\infty}{\longrightarrow} 0 \,.
\end{equation}
Such a limit is valid when $M > |\alpha_l| = r$, which gives us a condition on the radial coordinate $R < M/(2\sqrt{2}d)$.

Therefore, all terms with $q\neq 0$ are exponentially suppressed, and the leading order approximation consists in retaining only the $q=0$ contribution: 
\begin{equation}
    W(R, \phi) = (1-\Delta) \, \frac{e^{-R^2 - 2d^2}}{\pi N} I_0(r) + \Delta \, \frac{e^{-R^2 - d^2}}{\pi N} \sum_{\ell=0}^{M-1} e^{-d^2 e^{2i \frac{\pi}{M} \ell }} \, I_{0}(r e^{i \frac{\pi}{M} \ell}) \,.
\end{equation}
A further approximation can be performed on the sum over $\ell$. The larger contribution arises when the exponential is maximal, i.e., when
\begin{equation}
    e^{2i\pi/M\ell}=-1 \quad \Rightarrow \quad \ell = \frac{M}{2} \quad (\text{for even } M).
\end{equation}
We thus make the substitution $\ell = M/2+x$, which entails: 
\begin{equation}\label{SM:S}
    S \equiv \sum_{\ell=0}^{M-1} e^{-d^2 e^{2i \frac{\pi}{M} \ell }} \, I_{0}\left( r e^{i \frac{\pi}{M} \ell} \right) = \sum_{x=-M/2}^{M/2-1} e^{d^2 e^{i \frac{2\pi}{M} x }} \, I_{0}\left( i \,r \,e^{i \frac{\pi}{M} x} \right).
\end{equation}
Since $\left|{\rm exp}(d^2 e^{i \Theta x}) \right| = \left|{\rm exp}(d^2 \cos  (\Theta x)) \, {\rm exp}(i d^2 \sin  (\Theta x))\right| \leq e^{d^2}$, the largest contribution is tightly localized around $x=0$ and we can approximate the slowly varying modified Bessel function by its value at the maximum reached for $x=0$. In fact, from the Taylor expansion around $x=0$, we have:
\begin{equation}
    I_{0}( i \,r \,e^{i \frac{\pi}{M} x} ) \approx I_{0}( i \,r) + i \frac{\pi r x }{M} I_{1}( i \,r)  = J_{0}( r) + i \frac{\pi r x}{M}  J_{1}(r) \,,
\end{equation}
where we have used the properties of the Bessel function $I_0(ir)=J_0(r)$ and $I_1(ir)=J_1(r)$ valid for $r\in \mathbb{R}$. To neglect the contribution linear in $x$, we need that 
\begin{equation}
    \left|\frac{\pi r x }{M} \right| \ll \left|\frac{J_{0}( r)}{J_{1}( r)}\right| <1 \,,
\end{equation}
which is valid for $M \gg d$.

Hence, we can factorize the dominant contribution $J_0(r)$ out of the summation in \eqref{SM:S} such that
\begin{equation}
    S \approx J_0(r) \sum_{x=-M/2}^{M/2-1} e^{d^2 e^{i \frac{2\pi}{M} x }}.
\end{equation} 
The exponential inside the summation can be expanded as a power series, and we get: 
\begin{equation}
    S \approx  J_0(r) \sum_{x=-M/2}^{M/2-1} \sum_{n} \frac{ \left(d^2 e^{i \frac{2\pi}{M} x } \right)^n  }{n!}
    = J_0(r)  \sum_{n} \frac{d^{2n}}{n!} \sum_{x=-M/2}^{M/2-1} \left( e^{i \frac{2\pi }{M} n } \right)^x 
    = J_0(r)  \sum_{n} \frac{d^{2n}}{n!} M \delta_{n,qM} = J_0(r) M \sum_q \frac{d^{2qM}}{(qM)!} \,,
\end{equation}
where we have used the identity (\ref{eq:geometric_serie}). The summation $\sum_q \frac{d^{2qM}}{(qM)!}$ evaluated in $q=0$ gives exactly $1$, while the term with $q=1$ is equal to $d^{2M}/M!$. The latter is thus suppressed in the highly-packed limit $M \gg d^2$. Terms with $q$ larger than $1$ are even less relevant in the highly-packed limit.

Therefore, $S \approx J_0(r)M$ and the Wigner function simplifies to: 
\begin{equation}\label{app_eq:Wigner_bessel_zero_order}
    W(R, \phi) = W(R) = (1-\Delta)\,  \frac{e^{-R^2 - 2d^2}}{\pi N} I_0(r) + \Delta \, \frac{e^{-R^2 - d^2}}{\pi N } M J_{0}(r) = \frac{e^{-R^2-d^2}}{\pi N} \left[ (1-\Delta)\, e^{-d^2} I_0(r) + \Delta \, M \, J_{0}(r) \right].
\end{equation}

Solving $(1-\Delta)\, e^{-d^2} I_0(r) + \Delta \, M \, J_{0}(r)=0$ for $\Delta$ and minimizing over $r$ provides us the upper bound $\overline{\Delta}_c$ of $\Delta_c$ for $d \ge 1$:
\begin{equation}
    \overline{\Delta}_c = \min_{r}\frac{1}{1- e^{d^2}  M \frac{J_0(r)}{I_0(r)}} = \frac{1}{1 - e^{d^2}  M\frac{J_0(\pi)}{I_0(\pi)}} = \frac{1}{1+0.056 \,e^{d^2}  M},
\end{equation}
given that stationary points (i.e., where the first derivative vanishes) of the ratio $\frac{J_0(r)}{I_0(r)}$ occur at $r=0$, $r=\pm 3.19622$, $r=\pm 6.30644$, $r=\pm 9.43950 \dots$. As these values are well approximated by $r\cong n\pi$ with $n \in \mathbb{N}$, the minimum of $\frac{J_0(r)}{I_0(r)}$ is attained near $r=\pi$, for which $J_0(\pi)/I_0(\pi)\cong -0.056$.

In conclusion, if $\Delta \ge \overline{\Delta}_c$, then the Wigner function is surely negative.


\end{document}